\documentclass[%
 reprint,
 amsmath,amssymb,
 aps,
]{revtex4-2}

\usepackage{graphicx}
\usepackage{dcolumn}
\usepackage{bm}
\usepackage{amsmath}
\usepackage{endnotes}

\begin{document}

\preprint{APS/123-QED}

\title{Spin Polarization Control via Magnetic Field in Dissipative Bosonic Systems
}
\author{Yaoyuan Fan$^{1}$, Shuoyu Shi$^{2}$, Lang Cao$^{1}$, Qiuxin Zhang$^{3}$, Dong Hu$^{3}$, Yu Wang$^{3}$, Xiaoji Zhou$^{1}$$^,$$^{4}$}
\email{xjzhou@pku.edu.cn}
\affiliation{$^{1}$State Key Laboratory of Advanced Optical Communication Systems and Networks, School of Electronics, Peking University, Beijing 100871, China}
\affiliation{$^{2}$School of Physics, Peking University, Beijing 100871, China}
\affiliation{$^{3}$Changcheng Institute of Metrology and Measurement, Beijing 100095, China}
\affiliation{$^{4}$Institute of Carbon-Based Thin Film Electronics, Peking University, Shanxi, Taiyuan 030012, China}


\begin{abstract}

Engineering spin polarization in dissipative bosonic systems is crucial for advancing quantum technologies, especially for applications in quantum metrology and space-based quantum simulations. This work demonstrates precise magnetic moment control in multicomponent Bose gases during evaporative cooling via tailored magnetic fields. By adjusting the magnetic field gradients, null point position, and duration, we selectively tune evaporation rates of magnetic sublevels, achieving targeted spin polarization. Theoretical models, validated by numerical simulations and Stern-Gerlach experiments, reveal how magnetic fields reshape trapping potentials and spin-dependent dissipation. The results establish a dissipative spin-selection mechanism governing polarization evolution in evaporatively cooled Bose gases and provide a framework for engineering spin-polarized quantum states.

\end{abstract}

\maketitle


\section{\label{sec:level1}Introduction}

Dissipative bosonic systems, where energy and particle exchange with the environment play a critical role, have emerged as a pivotal platform for exploring non-equilibrium quantum phenomena \cite{Leggett2006, Bloch2008, Ritsch2013}. In such systems, the interplay between coherent dynamics and dissipation enables novel quantum states and phase transitions, offering unique opportunities for quantum simulation and precision measurement \cite{Barreiro2011, Devoret2013}. Among these systems, spinor Bose-Einstein condensates (BECs) exhibit rich spin dynamics due to their internal degree of freedom, where the spin components interact via both coherent collisions and external fields \cite{Loss1998, Cooper2019}. The ability to control spin dynamics through external parameters, such as magnetic fields, holds profound implications for quantum information processing \cite{DiVincenzo1998} and topological matter engineering \cite{Lin2009}.

Traditionally, studies of spin polarization control in BECs have focused on closed systems with conserved total atom number and magnetic moment \cite{Chang2004, Matuszewski2010, Horne2007}. Pioneering experiments by Chang et al. \cite{Chang2004} demonstrated dynamic evolution of spinor condensates under spin-orbit coupling and magnetic fields, revealing coherent spin-mixing oscillations and metastable states. Theoretical frameworks developed by Matuszewski \cite{Matuszewski2010} further elucidated the role of magnetic fields in stabilizing spin textures in trapped condensates. These works established a paradigm for understanding spin polarization control in equilibrium conditions.

However, the behavior of spinor BECs in dissipative environments, particularly during evaporative cooling---a non-equilibrium process essential for achieving quantum degeneracy---remains largely unexplored. Evaporative cooling inherently involves selective particle loss, where high-energy atoms escape the trap while low-energy atoms thermalize through collisions \cite{Chin2010}. In multi-component systems, this process becomes spin-dependent: magnetic fields modify the effective trapping potentials for different magnetic sublevels (\(m_F = \pm1, 0\)), thereby altering their evaporation rates and collision dynamics \cite{Petrich1995}. Recent studies by Horne et al. \cite{Horne2017} demonstrated magnetic field-assisted control of atomic trajectories in interferometry, hinting at the potential for spin-selective manipulation during cooling. Yet, a systematic investigation of how magnetic fields regulate the total magnetic moment through dissipative dynamics remains absent.

\begin{figure*}[htbp]
\includegraphics[width=\linewidth]{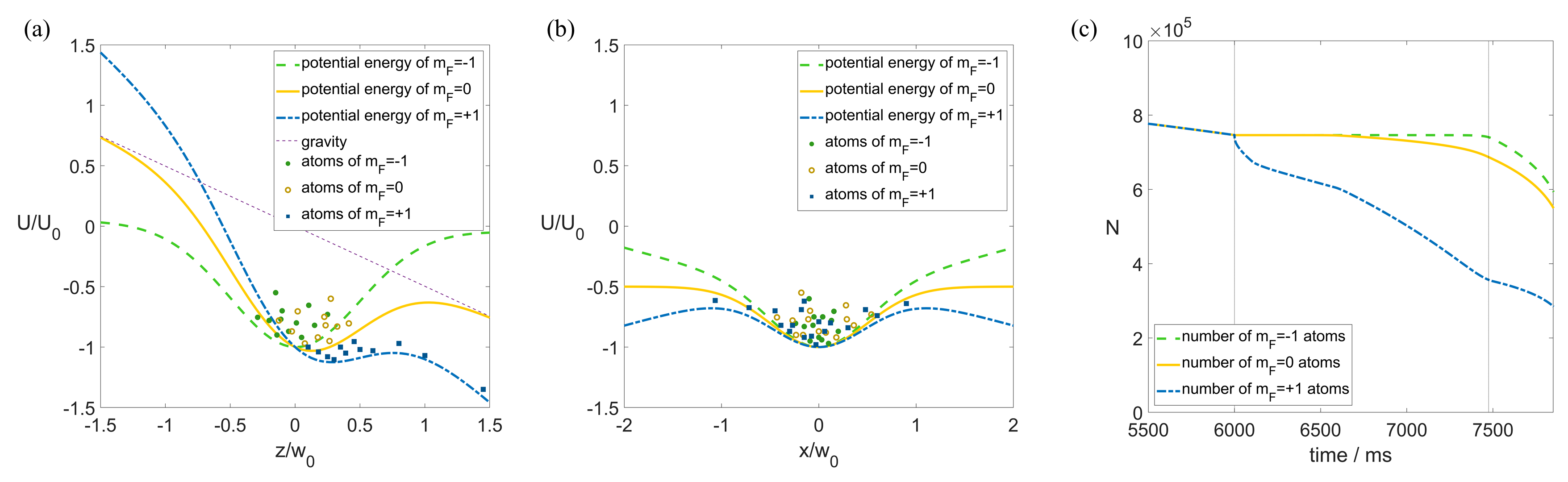} 
\caption{\label{fig:wide} (a)(b) Schematic representations of the potential configurations along the z- and x-directions under magnetic field conditions, when the quadrupole magnetic field zero point is above the optical trap. The purple short-dashed curves denote gravitational potentials. The yellow solid, blue dash-dotted, and green dashed curves respectively represent effective potentials experienced by atoms in magnetic sublevels 0, +1, and -1. The yellow hollow circles, blue squares, and green solid circles correspond to atoms in these three sublevels, each undergoing evaporation within their respective effective potential wells. (c) Numerical simulation of the time variation of numbers of atoms in different spin states. The two vertical lines represents opening and closing the magnetic field, respectively.
}
\end{figure*}

This work addresses this gap by investigating spin polarization control in a dissipative bosonic gas undergoing evaporative cooling under tunable magnetic fields. We demonstrate that the total magnetic moment can be precisely controlled by engineering the evaporation rates of spin components via magnetic field gradients and bias fields. Crucially, we uncover a cooperative cooling mechanism: atoms in higher magnetic sublevels, experiencing stronger gravitational tilts, exhibit enhanced collision rates that accelerate the cooling of lower sublevels. These findings are supported by a theoretical model combining dissipative spin-resolved evaporation equations with numerical simulations, validated through Stern-Gerlach experiments.

Our results not only advance the understanding of non-equilibrium spin dynamics but also provide a practical framework for preparing spin-polarized BECs---a critical capability for applications in quantum metrology \cite{Pezze2018} and synthetic gauge field engineering \cite{PhysRevLett.117.220401}. Furthermore, our methodology lays the groundwork for microgravity applications. By achieving precise control over spin-polarized states through tailored magnetic fields, we establish a framework for simulating microgravity-like conditions in ground-based experiments. Specifically, the ability to engineer spin-dependent potentials and compensate for gravitational tilts mimics the absence of gravity in space, enabling predictive modeling for future space-based quantum simulations \cite{Lundblad2019, PhysRevLett.123.160403}. This dual applicability underscores the versatility of our approach, where magnetic field-tuned spin dynamics serve as a universal tool for both terrestrial and extraterrestrial quantum engineering.

\section{\label{sec:level1}Theoretical model}

Evaporative cooling of trapped atoms is a typical example of dissipative bosonic system. We applied a magnetic field to the atoms, breaking the degeneracy of different magnetic sublevels, and tuned the evaporation rate of atoms with different spin states to investigate the spin dynamics in this system. In this section, firstly we describe the form of our trapping potential, then we extend the evaporation rate formula to a multicomponent gas and see how the magnetic quadrupole field deforms the trapping potential and thereby affects the evaporation rate of atoms with different spin states. In the end, we demonstrate the effect of cooperative cooling.

Our trap consists of three kinds of potential: gravity \( U_g = mgz \), optical dipole trap \( U_o \) induced by horizontal crossed red-detuned laser beams that overlap at their waists, and magnetic quadrupole trap \( U_m \), whose zero point can be displaced by a bias field:
\begin{equation}
U_o = U_\text{dip}I(\bm{r}) = U_\text{dip}\frac{2P}{\pi w_0^2}\left(e^{-2\frac{x^2+z^2}{w^2(y)}}+e^{-2\frac{y^2+z^2}{w^2(x)}}\right),
\end{equation}

\begin{equation}
U_m = \frac{\mu_\text{B}B'}{4}\sqrt{x^2+y^2+4(z-z_0)^2},
\end{equation}
where \( U_\text{dip} = -\frac{3\pi c^2}{2\omega_0^3}\left(\frac{\Gamma}{\omega_0-\omega}+\frac{\Gamma}{\omega_0+\omega}\right) \) is the strength coefficient of optical dipole trap\cite{GRIMM200095}, \(\omega_0\) is the resonance frequency, $\omega$ is the frequency of light, $\Gamma = \frac{\omega_0^3}{3\pi\varepsilon_0\hbar c^3}|\langle e|\mu|g\rangle|^2$ is  the damping rate corresponding to the spontaneous decay rate of the excited level, \( I(\bm{r}) \) is the intensity distribution of light, \( P \) is the power of each laser, \( w_0 \) is the beam waist, \( w^2(z) = w_0^2\left[1+\left(z\lambda/\pi w_0^2\right)^2\right] \) is the beam width, \( \mu_\text{B} \) is the Bohr magneton and \( B' \) is the magnetic field gradient. The effective potentials for different states of atoms are described in Fig. 1.

Both the center of the optical dipole trap and the zero point of the magnetic quadrupole trap are located on the z-axis. The optical potential forms a trap with finite depth that allows for evaporation, while gravity and the magnetic trap deform it. The magnetic field breaks the degeneracy of \( m_F = +1,0,-1 \) spin states, which makes the atom gas a spinful multicomponent gas. The total magnetic moment can be controlled via magnetic field. Roughly speaking, when the magnetic field zero point is above the well, the effective gravity of \( m_F = -1 \) is reduced and that of \( m_F = +1 \) is increased, making less \( m_F = -1 \) and more \( m_F = +1 \) atoms to escape, as shown in Fig. 1(a). When the zero point is below the well, the case for \( m_F = +1 \) and \( m_F = -1 \) will be reversed. Besides, the magnetic field also compress or expand the atomic clusters in the xy-plane, as shown in Fig. 1(b), which also changes the evaporation rate. To further investigate this phenomenon quantitatively, we extended the evaporation rate formula in \cite{PhysRevA.53.381, PhysRevA.55.1281} to our multicomponent gas.

By exploiting the Boltzmann equation of multicomponent gas, using the method in \cite{PhysRevA.53.381}, we can obtain the time derivative of the occupation number of atoms:
\begin{widetext}
\begin{equation}
\rho_i(\epsilon_4)\dot{f}_i(\epsilon_4)=\sum_j \frac{m\sigma}{\pi^2\hbar^3}\int\text{d}\epsilon_1\text{d}\epsilon_2\text{d}\epsilon_3\delta(\epsilon_1+\epsilon_2-\epsilon_3-\epsilon_4)\rho(\min[\epsilon_1,\epsilon_2,\epsilon_3,\epsilon_4])\left\{f_j(\epsilon_1)f_i(\epsilon_2)-f_j(\epsilon_3)f_i(\epsilon_4)\right\},\quad i,j=+1,0,-1
\end{equation}
\end{widetext}
where \( \sigma = 8\pi a^2 \) is the s-wave scattering cross-section. \( \epsilon_1 \) and \( \epsilon_2 \) are the total energies of the two atoms after collision, respectively. \( \epsilon_3 \) and \( \epsilon_4 \) are the total energies of the two atoms before collision, respectively. \( \rho(\epsilon) \) represents the density of states of particles in the energy range \( \epsilon \sim \epsilon + \text{d}\epsilon \), given by:
\begin{equation}
\rho(\epsilon) = \frac{2\pi (2m)^{3/2}}{(2\pi \hbar)^3} \int_{U(\bm{r}) \leq \epsilon} \text{d}^3 \bm{r} \sqrt{\epsilon - U(\bm{r})}.
\end{equation}
It is directly related to the shape of the trap. And \( f(\epsilon) \) is the occupation number:
\begin{equation}
f(\epsilon)=n_0\Lambda^3e^{-\epsilon/k_\text{B}T}\Theta(\epsilon_t-\epsilon)
\end{equation}
where \( \Theta \) is the Heaviside step function, \( \epsilon_t \) is the truncated energy of evaporative cooling, \( \Lambda = \sqrt{\frac{2\pi\hbar^2}{mk_\text{B}T}} \)and \( n_0\Lambda^3 \) represents the occupation number of the lowest energy state.

Here we use Boltzmann distribution to approximate the Bose-Einstein distribution to simplify calculations, which is reasonable because during the time when the magnetic field is turned on (see Fig. 3(b)), the Bose gas cluster is far from condensation.

By detailed balance, when the trapping potential decreases very slowly, the evaporation rate is approximately the collision rate of atoms with energy greater than the trap depth\cite{WOS:A1996BH82Z00004}:
\begin{equation}
\dot{N}_{\text{ev},i}=-\int_{\epsilon_{t,i}}^{\infty}\text{d}\epsilon_4\rho_i(\epsilon_4)\dot{f}_i(\epsilon_4)
\end{equation}

To determine \( \min[\epsilon_1,\epsilon_2,\epsilon_3,\epsilon_4] \), we need to use an approximation that when the trapping potential changes very slowly, most of the atoms are in the energy range \( \epsilon < \epsilon_{t,\text{min}} \), \( \epsilon_{t,\text{min}} \) is the lowest potential depth among three states. So \( \epsilon_4 > \epsilon_{t,i} \geq  \epsilon_{t,\text{min}} > \epsilon_1,\epsilon_2 \). And \( \epsilon_1 + \epsilon_2 = \epsilon_3 + \epsilon_4 \) for elastic collision, so \( \min[\epsilon_1,\epsilon_2,\epsilon_3,\epsilon_4] = \epsilon_3 \). By substituting equations (3),(5) into (6), we obtain:
\begin{widetext}
\begin{equation}
\dot{N}_{\text{ev},i}=-\sum_j\frac{m\sigma}{\pi^2\hbar^3}\int_0^{\epsilon_{t,j}}\text{d}\epsilon_3\int_{\epsilon_3}^{\epsilon_{t,j}}\text{d}\epsilon_1\int_{\epsilon_3+\epsilon_{t,i}-\epsilon_1}^{\epsilon_{t,i}}\text{d}\epsilon_2\rho_j(\epsilon_3)f_j(\epsilon_1)f_i(\epsilon_2)=-\sum_j N_iN_j\sigma\bar{v} e^{-\eta_i}\frac{V_{\text{ev},j}}{V_{\text{e},i}V_{\text{e},j}},\quad i,j=+1,0,-1
\end{equation}
\end{widetext}

where \( \bar{v} = \sqrt{8k_\text{B}T/\pi m} \) is the average velocity, \(\eta \equiv \epsilon_t / k_\text{B} T\) is the truncation parameter, \(V_\text{e} \equiv N/n_0 \) is the reference volume (\(   n_0 \) is the reference density\cite{PhysRevA.53.381}), and \(V_{\text{ev}}\) is the effective volume for elastic collisions. These parameters constitute the connection between the trap properties and the evaporation rate. \(\eta\) represents the trap depth. \( V_{\text{e}} \) and \( V_{\text{ev}} \) are given by following expressions:
\begin{equation}
V_{\text{e}} = \frac{\Lambda^3}{(2\pi\hbar)^3} \int_0^{\epsilon_t} \rho(\epsilon) e^{-\epsilon / k_\text{B} T} \text{d}\epsilon,
\end{equation}
\begin{equation}
V_{\text{ev}} = \Lambda^3 \int_0^{\epsilon_t} \rho(\epsilon) \left[ \left( \eta - \frac{\epsilon}{k_\text{B} T} - 1 \right) e^{-\epsilon / k_\text{B} T} + e^{-\eta} \right] \text{d}\epsilon.
\end{equation}

\begin{figure}[!b]
    \centering
    \includegraphics[width=\linewidth]{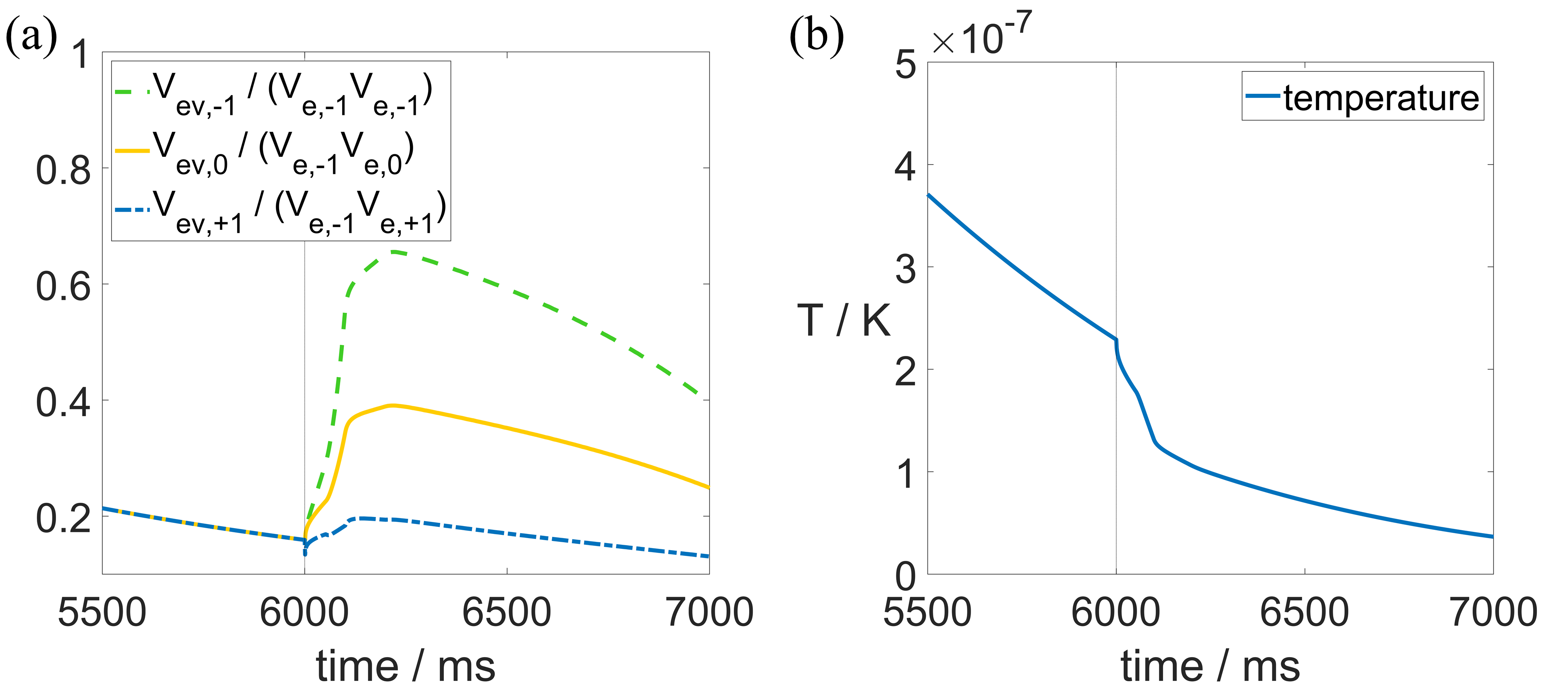}
    \caption{\label{fig:wide} (a) The numerical result of relative value of the volume fraction \( V_{\text{ev},j}/V_{\text{e},-1}V_{\text{e},j} \) over time. The vertical line represents the opening time of magnetic field. The magnetic field enlarges the volume fraction in each collision term of formula (7) of \( m_F = -1 \) atoms. (b) The numerical result of the temperature over time. The vertical line represents the opening time of magnetic field. It is obvious that the magnetic field accelerates the cooling process. 
}
    \label{fig:Fig2}
\end{figure}

From the evaporation rate formula (7), we can find that the truncation parameter \( \eta_i \) only appears in the formula of \( m_F = i \) state, and the differences between \(\eta_{-1}\), \(\eta_{0}\) and \(\eta_{+1}\) will lead to different evaporation rates of different spin states, and therefore the total magnetic moment can be tuned. However, the reference density \( n_{0,i} = N_i/V_{\text{e},i} \) and the effective volume for elastic collision \( V_{\text{ev},i} \) for each spin state are shared by all three spin states, which means that the changes in these two parameters will influence the overall evaporation rate, enhancing the cooperative cooling effect. Thus, we need to explore the influence of magnetic field on these parameters.

We first focus on the truncation parameter \( \eta \equiv \epsilon_t/k_\text{B}T \). When the magnetic field is turned on, \( m_F = -1 \) atoms will be pulled towards its zero point, while \( m_F = +1 \) atoms will be repelled from its zero point. So when the zero point is above the well, the effective gravity of \( m_F = -1 \) is reduced and that of \( m_F = +1 \) is increased, making the effective trap depth \( \epsilon_t \) of \( m_F = -1 \) higher and \( m_F = +1 \) lower. This difference in trap depth will be reflected in the exponential term \( e^{-\eta} \) in the evaporation rate formula, allowing more \( m_F = +1 \) and fewer \( m_F = -1 \) atoms to escape. In this case, the massive evaporation of \( m_F = +1 \) atoms accelerates the cooling process, as shown in Fig. 2(b), and preserves more \( m_F = -1 \) atoms, as shown in Fig. 1(c). When the zero point is below the well, the situation will reverse.

Secondly, the magnetic field applies horizontal centripetal and centrifugal forces to the \( m_F = -1 \) and \( m_F = +1 \) states respectively, increasing the trap frequency of \( m_F = -1 \) spin state and lowering that of \( m_F = +1 \) spin state, which changes the density of state \( \rho(\epsilon) \) of atoms and thereby changes the \( V_\text{ev} \) and \( V_\text{e} \) of both states. From our numerical result, when the magnetic field is turned on, the volume fraction \( V_{\text{ev},j}/V_{\text{e},i}V_{\text{e},j} \) in each term of the evaporation rate formula (7) of \(m_F = -1\) state will increase, as shown in Fig. 2(a), and that of \(m_F = +1\) state will decrease. It will increase the total magnetic moment. When the magnetic field zero point is near the potential well, the horizontal centripetal and centrifugal forces are stronger, and this effect is more significant.

The cross terms in the evaporation rate formula reveal the kinetics of cooperative cooling. An atom in a certain state collides with another atom, which is in the same state or in the other two states, so the evaporation rate is proportional to the product of the densities of two states \( N_iN_j/V_{\text{e},i}V_{\text{e},j} = n_{0,i}n_{0,j} \), and the effective volume of elastic collision \( V_{\text{ev},j} \) is of the atom that collides with it. In addition, different states share the same \( \bar{v} \) and \( T \). These terms show the influence of one state on the other, which is the basis of cooperative cooling. The truncation parameter \( \eta_i \) of a certain state appears only in the evaporation formula of itself, resulting in a rich spin dynamics in the following text.

\section{\label{sec:level1}Experiment Set-up}

\begin{figure*}[htbp]

\includegraphics[scale=0.41]{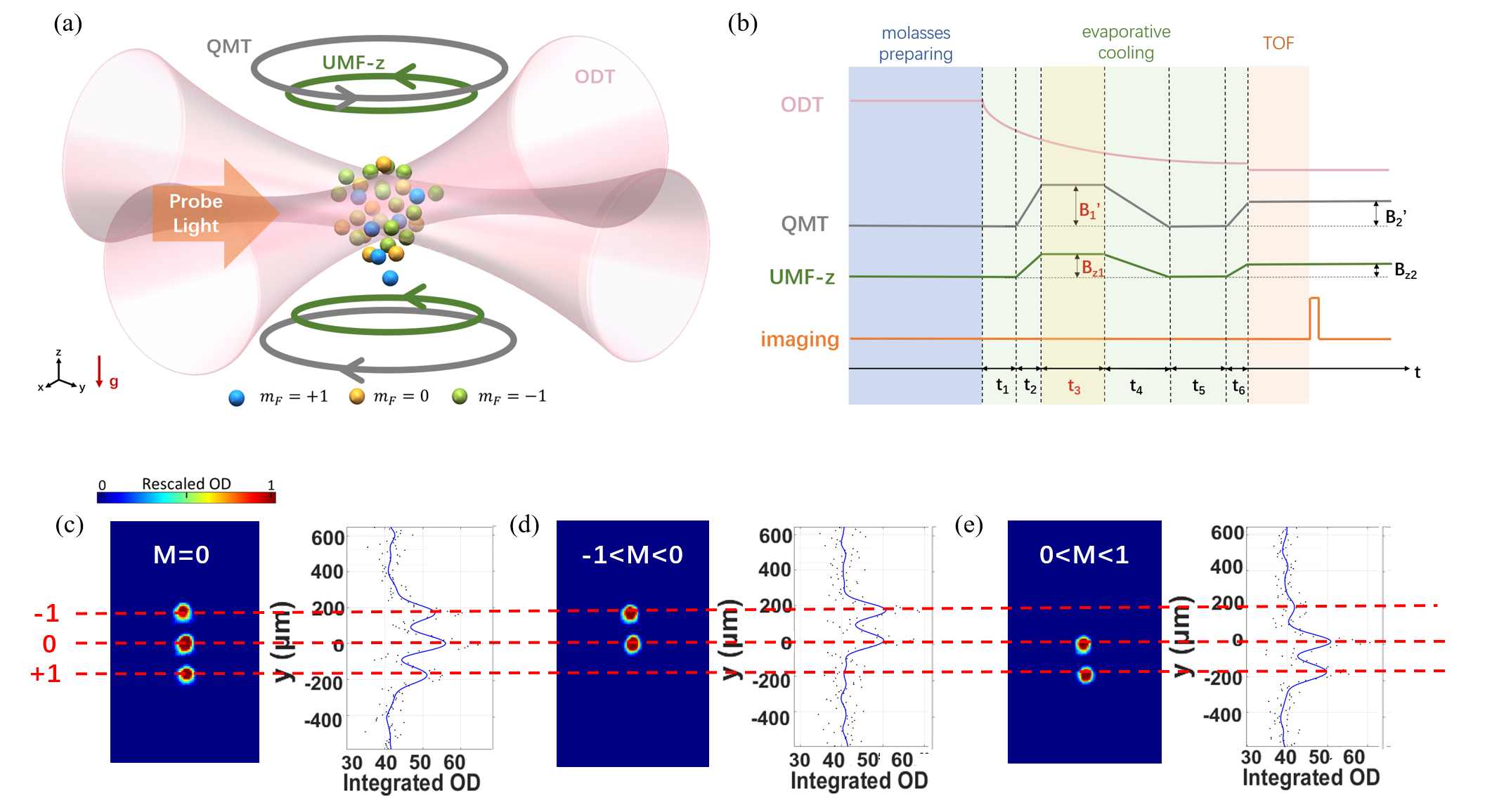} 
\caption{\label{fig:epsart}(a) Schematic diagram of the experimental setup. Atoms are confined in a pair of mutually orthogonal crossed optical dipole traps (ODTs) within the horizontal plane. Two sets of coils along the $z$-direction generate a quadrupole magnetic trap (QMT) and a uniform magnetic field along $z$-axis (UMF-$z$), respectively. The probe beam propagates along the bisecting direction of the crossed optical traps in the horizontal plane, enabling observation of atomic cloud separation after the Stern-Gerlach experiment. Distinct colors represent atoms in different magnetic sublevels. 
(b) Experimental timing sequence. The protocol comprises three stages: Molasses preparation, evaporative cooling, and time-of-flight (SG experiment). During the evaporative cooling phase, magnetic fields are adiabatically applied through linear ramping and shutdown. Key experimental parameters under investigation include: magnetic field duration $t_3$ (magnetic field manipulation time), magnetic field gradient magnitude during this period (affecting atomic forces), and bias field strength (determining relative position between magnetic field and atoms).
(c)-(e) Absorption images demonstrating spatial segregation along gravitational direction in the Stern-Gerlach experiment under degenerate magnetic sublevels. (c) The total magnetic moment approximates zero.
(d) Predominance of $-1$ magnetic sublevel atoms. Total magnetic moment $< 0$. (e) Dominance of $+1$ magnetic sublevel atoms.
Total magnetic moment $> 0$.
}

\end{figure*}

To validate the theoretical model of spin-dependent evaporation dynamics, we developed a hybrid trapping system integrating optical and magnetic confinement. Three orthogonal pairs of compensation coils (x, y, z) were employed to ensure the quadrupole magnetic field axis was precisely aligned with both the gravitational direction and the optical trap center. This setup enables quantitative control of the bias field (\(B_{z1}\)), the magnetic field duration ($t_3$) and the quadrupole magnetic gradient (\(B_1'\)) as three tuning parameters.

Fig. 3(a) schematically illustrates our hybrid trapping configuration. The ultracold atomic ensemble is confined at the beam waist intersection of two horizontally crossed optical dipole traps (ODTs), integrated with a quadrupole magnetic trap (QMT) and a z-axis uniform magnetic field (UMF-z). For clarity and to emphasize key features, the three pairs of compensation coils have been omitted from the diagram. Through active positioning of the optical traps, we establish a precise vertical alignment where the magnetic field zero-point of the QMT resides below the optical confinement center when UMF-z is disabled. By strategically manipulating both the optical trap position and UMF-z strength, we achieve full control over the magnetic null point's vertical displacement relative to the trapping region. This engineered configuration allows selective positioning of the magnetic zero point either above or below the optical trap through coordinated adjustments of these parameters.

The timing sequence is shown in Fig. 3(b). The experimental sequence comprises three primary stages. First is optical molasses preparation (blue-shaded region): During this phase, we prepare \(^{87}\text{Rb}\) atoms in the \( F=1 \) hyperfine ground state, establishing the initial conditions for spin-dependent dynamical control. Following the optical molasses preparation, the experiment progresses to the optical evaporative cooling phase (green background), where we implement magnetic field regulation protocols to tune the system's spin-state dynamics; after this stage, the atomic ensemble undergoes free evolution under combined quadrupole magnetic and gravitational fields, followed by absorption imaging with Stern-Gerlach separation to resolve the spin population distribution.

The evaporative cooling stage constitutes the critical phase for magnetic field manipulation in our experiment, spanning a total duration of $\mathrm{9~s}$. At $t_1$ into this cooling process, we implement linear activation protocols for both the QMT and UMF-z, achieving full operational strength over $t_2 = \mathrm{100~ms}$ through precisely controlled current ramping. This procedure ensures that the magnetic field zero point does not shift during the application of the magnetic fields. Once the magnetic fields are turned on, they
are maintained for a tunable duration of $t_3$, which is designated as one of three key control variables in our experimental protocol, governs the magnetic field-induced differential evaporation rates between magnetic sublevels during this critical temporal window. This precisely adjustable interval ($0~\mathrm{s} \leq t_3 \leq 6~\mathrm{s}$) serves as the primary driver for the spin-state dynamics evolution demonstrated in Section~II, where the controlled variation of $t_3$ enables systematic investigation of magnetic-sublevel-dependent evaporation processes. After this period, the two magnetic fields are linearly turned off over a time interval of $t_4$, which is experimentally fixed at $\mathrm{1300~ms}$ based on empirical optimization of system stability.

Two additional adjustable parameters beyond $t_3$ govern the system's behavior within the magnetic field implementation sequence: the quadrupole field gradient $B_1'$ and uniform axial field strength $B_{z1}$. The gradient force experienced by the $^{87}\text{Rb}$ ($F=1$) atomic cloud scales linearly with $B_1'$ -- specifically, at $B_1' = 30.5~\mathrm{Gs/cm}$ the magnetic force precisely counterbalances gravitational acceleration ($F_\text{mag} = m \text{g}$). Our experimental configuration positions the magnetic null point $\Delta z = \mathrm{-0.62~mm}$ below the optical trap when $B_{z1}=0$. The null point displacement under finite $B_{z1}$ follows:
\begin{equation}
\label{eq:10}
\Delta z = -0.62~\mathrm{mm} + \frac{B_{z1}}{B_1'} 
\end{equation}
This parametric coupling enables precise spatial control of the magnetic zero point relative to the optical confinement center through coordinated adjustments of $B_1'$ and $B_{z1}$. In the experiment, 
$B_{z1}$ is adjustable within the range of $0$ to $5~\mathrm{Gs}$. Therefore, even under a gradient of $30.5 \mathrm{Gs/cm}$, the magnetic field zero point can be moved up and down through the center of the optical trap by adjusting $B_{z1}$.

It is noted that after the magnetic field is turned off at time $t_5$, the quadrupole magnetic field and the uniform magnetic field in the vertical direction are again linearly ramped up simultaneously and held at a constant value during the time-of-flight (TOF) stage. This procedure is designed for the Stern-Gerlach experiment, as the atoms in the three magnetic sublevels experience different forces in the vertical direction. This method allows spatial separation of the three components, facilitating the investigation of atomic population distribution and the total magnetic moment of the system. The linear ramp-up is also chosen to avoid heating caused by abrupt shifts in the magnetic field zero point.

The investigation of atomic spin polarization control in the experiment is primarily manifested through the total spin magnetic moment \( M \) of the atomic ensemble. \( M \) can be quantitatively resolved via absorption imaging as shown in Fig. 3(c)--(e). In horizontally acquired absorption images, the spatial distribution reveals distinct magnetic sublevel populations: atoms in the \( m_F = -1 \) state occupy the uppermost region due to the compensation effect of the magnetic force on gravity, those in the \( m_F = 0 \) state reside in the central zone, and those in the \( m_F = +1 \) state are localized in the lowermost region due to the superposition effect of the magnetic force on gravity. The population of atoms in each magnetic sublevel is determined by integrating the optical depth across the corresponding spatial domains. Based on the definition of magnetic moment, we can calculate:
\begin{equation}
\label{eq:11}
M = \frac{N_+ - N_-}{N_+ + N_- + N_0}
\end{equation}
where \( N_+ \), \( N_- \), and \( N_0 \) represent the number of atoms in each respective spin state.

\section{Experimental demonstration of spin polarization control}

\begin{figure*}[htbp]
\includegraphics[scale=0.33]{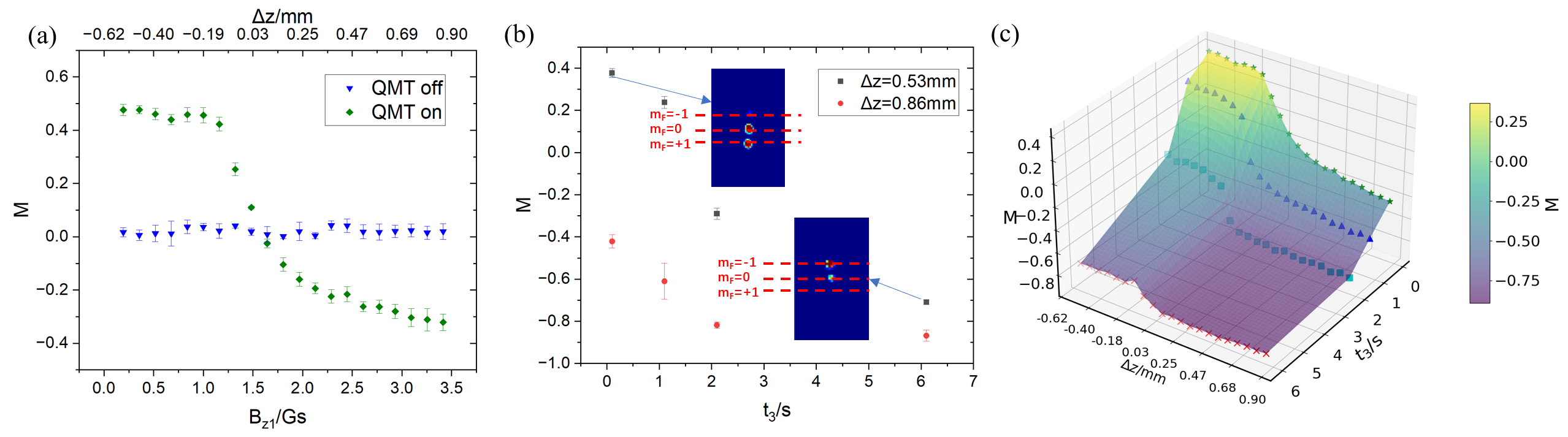} 
\caption{\label{fig_4} (a) (b) Variation of the system's magnetic moment as a function of the magnitude of UMF-z $B_{z1}$ (a) and the magnetic field duration $t_3$ (b) under a fixed magnetic field gradient. (c) Surface plot of the system's total spin magnetic moment \( M \) as a function of the bias magnetic field strength \( B_{z1} \) and the magnetic field turn-on duration \( t_3 \). The four curves in the surface plot correspond to the data points measured at fixed values of \( t_3 \) with varying \( B_{z1} \). Lighter colors in the surface plot represent larger values of the system's total magnetic moment. }
\end{figure*}

Different experimental parameters affect the forces acting on the atoms. As described in the previous model of the tilted potential well, different spin states exhibit varying degrees of tilt, with states that have smaller tilts being more readily selected. By adjusting the experimental parameters, we can effectively control the proportion of spin states, thereby modulating the overall magnetization.

This section systematically examines the effects of three parameters in experimental timing sequence Fig. 3(b) -- the magnitude of UMF-z $B_{z1}$ (determining relative position between magnetic field and atoms), the magnetic field duration $t_3$ (manipulation time), and the gradient magnitude of QMT $B_1'$ -- on the total magnetic moment of the system. In the following, we present experimental results demonstrating the effect of these parameter variations on total magnetization. By using absorption imaging, we obtain the atomic count, allowing us to calculate the overall magnetization based on the distribution of different spin states.

\subsection{$B_{z1}$ Dependence}

Prior to analyzing other system parameters, precise determination of the magnetic null point position is critical as it governs the force polarity experienced by different magnetic sublevels. Following Equation~(\ref{eq:10}), we achieve z-axis null point control through $B_{z1}$ adjustments, with 23~Gs/cm gradient corresponding to $\Delta z = 0.45~\mathrm{ mm/Gs}$ displacement sensitivity. Crucially, our experimental protocol implements geomagnetic compensation via x/y-axis coil current optimization, ensuring the null point of the QMT lies on the same vertical line as the center of the ODTs. 

We first investigated the influence of magnetic null point position on the system's total magnetic moment during evaporation. The experiment was conducted with fixed parameters $B_1' = 23~\mathrm{Gs/cm}$ and $t_3 = 100~\mathrm{ms}$, while varying $B_{z1}$ from $0$ to $3.5~\mathrm{Gs}$. According to Equation~(\ref{eq:10}), this parameter range translates to a null point displacement from approximately $0.62~\mathrm{mm}$ below to $0.90~\mathrm{mm}$ above the optical trap center. For each null point configuration, we analyzed the magnetic moment distribution through Stern-Gerlach separation, with the statistically averaged results presented in Fig.~\ref{fig_4}(a) based on multiple experimental realizations.

In Fig.~\ref{fig_4}(a), the blue inverted triangles represent control measurements with disabled quadrupole magnetic field (QMT off). In this configuration, varying $B_{z1}$ induces no change in the system's net magnetic moment (persistent zero value), as all three magnetic sublevels ($m_F = -1, 0, +1$) remain degenerate without magnetic gradient-induced state splitting.

When activating the quadrupole field ($B_1' = \mathrm{23~Gs/cm}$), the green diamond markers demonstrate null point position-dependent magnetic moment modulation. Each datum corresponds to a specific $B_{z1}$ setting that vertically displaces the magnetic null point according to $\Delta z = B_{z1}/B_1'$, thereby breaking the spin-state degeneracy and generating measurable magnetization through spin-selective evaporation.

When the magnetic null point resides below the optical trap ($B_{z1} < 1.35~\mathrm{Gs}$), experimental observations reveal a persistent positive net magnetic moment, indicative of $m_F=+1$ atoms undergoing magnetic levitation that counteracts gravitational sag. This aligns with our theoretical framework where $m_F=-1$ atoms experience accelerated evaporation due to trap potential tilting, synergistically cooling the $m_F=0$ and $+1$ populations through rethermalization collisions -- ultimately enhancing the $m_F=+1$ fraction compared to equilibrium predictions.

As $B_{z1}$ increases beyond $1.35~\mathrm{Gs}$ (null point displacement $\Delta z > 0$), $M$ undergoes nonmonotonic collapse, decreasing by about $7\%$ per $0.1~\mathrm{Gs}$ increment until transitioning to negative values when $\Delta z > 0~\mathrm{mm}$. This phase inversion originates from reversed magnetic gradient polarity relative to gravity, where $m_F=-1$ atoms now occupy the levitated metastable state while $m_F=+1$ components are preferentially expelled from the trap.

\subsection{$t_3$ Dependence}

Apart from the fact that the position of the magnetic field zero point can intuitively alter the system's total magnetic moment, the experiment also revealed that changing the magnetic field duration \(t_3\) has an interesting effect on the total magnetic moment, and this effect is not as straightforward.

In Fig.~\ref{fig_4}(b), the black square markers represent the case where the magnetic field is located below the optical trap, while the red circular markers correspond to the situation where the magnetic field is above the optical trap. It can be seen that as \(t_3\) increases, the total magnetic moment of the system decreases regardless of whether the magnetic field zero point is above or below the optical trap. In fact, when the magnetic field zero point is below the optical trap, increasing \(t_3\) beyond a certain value causes \(M\) to change from positive to negative.

To thoroughly investigate the dependence of \(M\) on different magnetic field application times \(t_3\) and magnetic null point positions \(\Delta z\), we varied the magnetic null point position under four different values of \(t_3\): 0.1 s, 1.1 s, 2.1 s and 6.1 s. Based on the results, we plotted the surface shown in Fig.~\ref{fig_4}(c). It can be observed that a higher magnetic null point position and a longer magnetic field application time both lead to a shift of the total magnetic moment \(M\) toward the negative direction. This phenomenon can be well explained using the theoretical model presented earlier, in conjunction with the potential well structure of the \(\pm1\) magnetic sublevels in the quadrupole magnetic trap.

Regardless of the bias magnetic field strength, a longer levitation duration causes the total magnetization to shift further toward negative values. This is because the \( -1 \) state is a low-field-seeking state, whereas the \( +1 \) state is a high-field-seeking state. With extended confinement in the trap, the \( +1 \) state is more likely to escape from the trap center, while the \( -1 \) state is preferentially selected, resulting in a shift in magnetization toward negative values.

\subsection{\( B_1' \) Dependence}

\begin{figure}[htbp]
\includegraphics[scale=0.25]{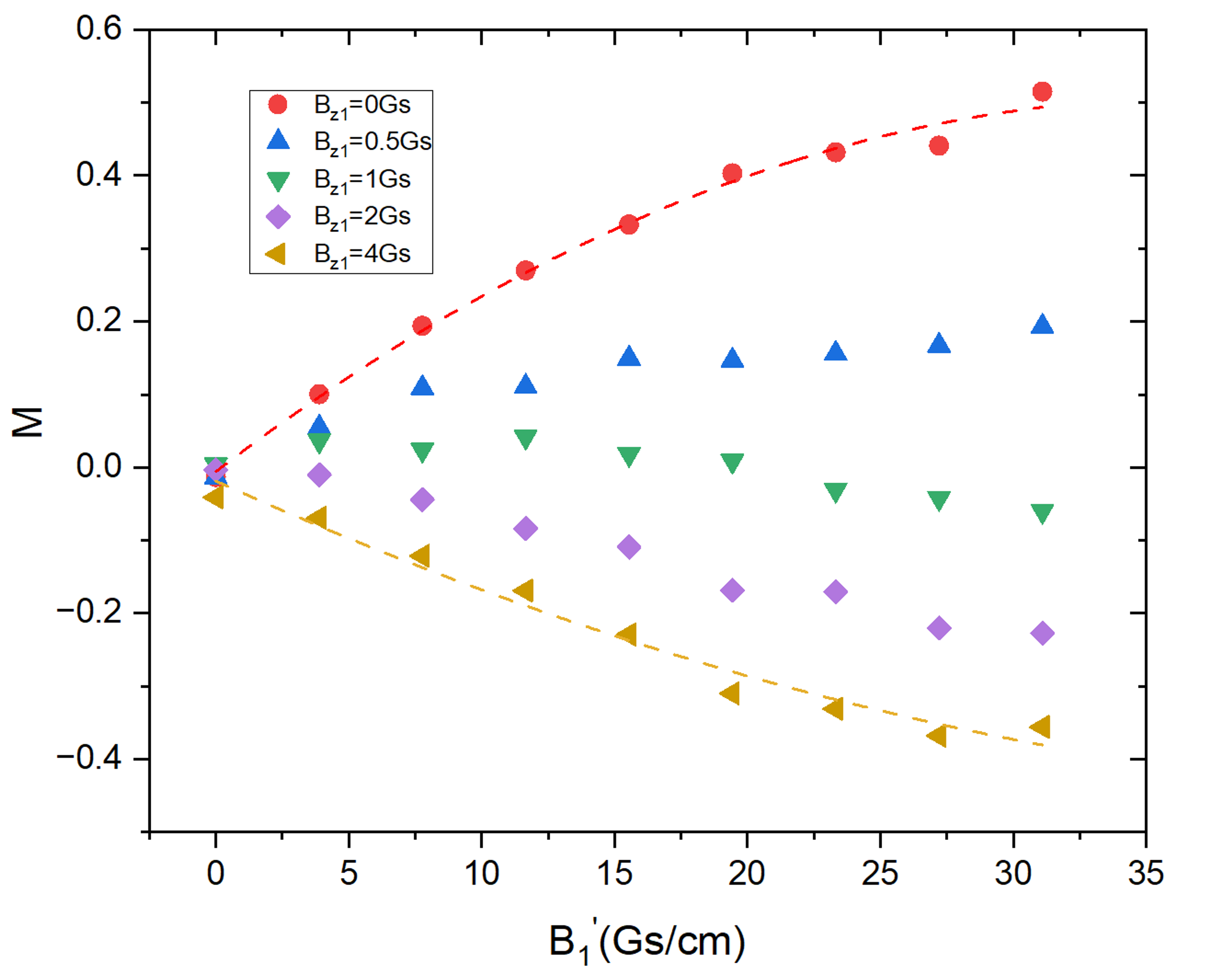} 
\caption{\label{fig_5}Experimental data of the system's magnetic moment as a function of magnetic field gradient for different bias magnetic field strengths. The horizontal axis represents the quadrupole magnetic field gradient, and the vertical axis represents the total magnetization.}
\end{figure}

Another intuitive factor affecting the size of the magnetic moment is the magnetic field gradient \( B_1' \). Fig.~\ref{fig_5} presents experimental data showing how the magnetic moment varies with the magnetic field gradient under different \( B_1' \). 

It can be seen that when the magnetic field gradient is zero, the magnetic moment is also zero, regardless of the strength of the bias magnetic field, indicating that the population of the three magnetic sublevels is equal. As the magnetic field gradient gradually increases, the bias magnetic field starts to play a role. When the bias magnetic field is selected such that the magnetic field null point is located below the atoms, the larger the magnetic field gradient, the larger the total magnetic moment of the system after evaporation, where most of the atoms are concentrated in the $+1$ state. Conversely, when the bias magnetic field is chosen such that the magnetic field null point is above the atoms, the larger the magnetic field gradient, the smaller the total magnetic moment of the system after evaporation, where most of the atoms are concentrated in the $-1$ state.

An important observation in Fig.~\ref{fig_5} is that, for $B_{z1}=0$ Gs and $B_{z1}=4$ Gs, as the magnetic field gradient increases, the variation of the magnetic moment follows a parabolic trend. This is consistent with the model described in the theoretical section. 

Taking the yellow inverted triangular experimental data points below in Fig.~\ref{fig_5} as an example. When the magnetic trap null point is above the optical trap, the gravity of the $-1$ state atoms is compensated, while the gravity of the $+1$ state atoms is enhanced, resulting in a negative total magnetic moment. As the magnetic field gradient increases, the quadrupole zero point becomes closer to the potential trap, which makes the horizontal centripetal and centrifugal forces much stronger, and thereby increases the evaporation rate of atoms in \( m_F = \pm 1 \) states, as discussed in part two, leading to a reduction in atomic density and a further slowdown of evaporation. Correspondingly, due to the cooperative effects, the evaporation rate of the $+1$ state increases, and the magnetic moment decreases as the zero point moves upward.

The regulation of dissipative bosonic systems by the magnetic field is reflected not only in the total magnetic moment of the system but also in its effect on the evaporation rate. As a result of these differences in $\eta$, atoms in higher magnetic sublevels undergo elastic collisions at a higher rate compared to those in lower sublevels. These more frequent collisions facilitate energy transfer from the higher energy atoms to the lower energy ones, enhancing the evaporative cooling of the lower sublevel components. This mechanism creates a cooperative cooling effect, where the faster-moving atoms in higher sublevels accelerate the evaporation of the slower-moving atoms in the lower sublevels. The enhanced collision rate in the higher sublevels promotes a more efficient cooling process for the entire multicomponent gas.

The cooperative nature of this process significantly improves the overall efficiency of evaporative cooling in the multicomponent system. By increasing the rate at which energy is transferred between components, the cooling process becomes more effective than in single-component systems. This interaction not only leads to a more efficient cooling mechanism but also opens up new possibilities for controlling the cooling dynamics through manipulation of magnetic field parameters. This cooperative cooling effect could have important implications for future precision measurements and quantum simulations involving multicomponent Bose gases.

\section{\label{sec:level1}Discussions and Conclusions}
According to our theoretical model and experiment results, three methods for obtaining spin-pure states via evaporative cooling can be proposed. First, to obtain a pure \( m_F = -1 \) state, the magnetic quadrupole trap zero point needs to be adjusted to a position far above the optical trap. The gravitational compensation effect of the magnetic quadrupole trap should be utilized, and the magnetic quadrupole trap must be applied over an extended period. At the end of this period, the intensity of the optical trap should be slightly less than the compensating gravitational force, ensuring that atoms in the \( m_F = +1 \) and \( m_F = 0 \) states evaporate out of the system along the \( z \)-direction.

Secondly, to obtain a pure \( m_F = +1 \) state, the method is essentially the same as that for the \( m_F = -1 \) state, except that the magnetic quadrupole trap zero point needs to be adjusted to a position far below the optical trap.

Finally, to obtain a pure \( m_F = 0 \) state, a two-stage process with different zero point positions of the magnetic quadrupole trap is required. In the early evaporation stage, when the intensity of the optical trap is relatively high, a magnetic quadrupole trap with its zero point at the center of the optical trap should be applied to best utilize its centrifugal force on \( m_F = +1 \) state, allowing it to evaporate along the laser beam. Then, in the later stage of evaporation (when the optical trap must be strong enough to compensate for gravity, otherwise the atoms in the \( m_F = 0 \) state will evaporate completely), a magnetic quadrupole trap with its zero point below the optical trap should be applied, allowing atoms in the \( m_F = -1 \) state to evaporate.

In conclusion, this study presents a detailed exploration of the evaporative cooling process in multicomponent Bose gases, focusing on the spin polarization control and the effects of differential gravitational forces on the magnetic sublevels. By compensating for gravity with quadrupole magnetic fields, we were able to break the degeneracy of the magnetic sublevels and create a three-component Bose gas. Our results reveal that the different magnetic sublevels, due to their varying effective gravitational influences, lead to differential collision rates during the evaporative cooling process. Atoms in higher magnetic sublevels, experiencing greater tilt in their potential wells, undergo more frequent elastic collisions, which in turn facilitate the cooling of lower energy components through cooperative interactions.

The theoretical model we developed successfully explains how these interactions influence the evaporation rate and the final total magnetization of the gas. Experimental results corroborate our model, showing a high degree of agreement between theory and experiment. The cooperative cooling effect observed not only enhances the efficiency of the evaporative cooling process but also provides a novel method for controlling the final total magnetization by tuning the magnetic field parameters.

This work contributes to a deeper understanding of the dynamics in multicomponent Bose gases and offers new insights into how the interplay between different magnetic sublevels can be exploited to control the cooling process. These findings have potential applications in a variety of quantum experiments, such as precision measurements and quantum simulations\cite{Moan2020, Beydler2023}, where the ability to manipulate the total magnetization and cooling efficiency could be crucial.

\begin{acknowledgments}
This work was supported by the National Key Research and Development Program of China (Grant Nos. 2021YFA0718300 and 2021YFA1400900), the National Natural Science
 Foundation of China (Grant Nos. 11920101004, 11934002, and 92365208), and the Space Application System of China Manned Space Program.
\end{acknowledgments}

\nocite{*}

\bibliography{apssamp}

\begin{thebibliography}{25}%
\makeatletter
\providecommand \@ifxundefined [1]{%
 \@ifx{#1\undefined}
}%
\providecommand \@ifnum [1]{%
 \ifnum #1\expandafter \@firstoftwo
 \else \expandafter \@secondoftwo
 \fi
}%
\providecommand \@ifx [1]{%
 \ifx #1\expandafter \@firstoftwo
 \else \expandafter \@secondoftwo
 \fi
}%
\providecommand \natexlab [1]{#1}%
\providecommand \enquote  [1]{``#1''}%
\providecommand \bibnamefont  [1]{#1}%
\providecommand \bibfnamefont [1]{#1}%
\providecommand \citenamefont [1]{#1}%
\providecommand \href@noop [0]{\@secondoftwo}%
\providecommand \href [0]{\begingroup \@sanitize@url \@href}%
\providecommand \@href[1]{\@@startlink{#1}\@@href}%
\providecommand \@@href[1]{\endgroup#1\@@endlink}%
\providecommand \@sanitize@url [0]{\catcode `\\12\catcode `\$12\catcode `\&12\catcode `\#12\catcode `\^12\catcode `\_12\catcode `\%12\relax}%
\providecommand \@@startlink[1]{}%
\providecommand \@@endlink[0]{}%
\providecommand \url  [0]{\begingroup\@sanitize@url \@url }%
\providecommand \@url [1]{\endgroup\@href {#1}{\urlprefix }}%
\providecommand \urlprefix  [0]{URL }%
\providecommand \Eprint [0]{\href }%
\providecommand \doibase [0]{https://doi.org/}%
\providecommand \selectlanguage [0]{\@gobble}%
\providecommand \bibinfo  [0]{\@secondoftwo}%
\providecommand \bibfield  [0]{\@secondoftwo}%
\providecommand \translation [1]{[#1]}%
\providecommand \BibitemOpen [0]{}%
\providecommand \bibitemStop [0]{}%
\providecommand \bibitemNoStop [0]{.\EOS\space}%
\providecommand \EOS [0]{\spacefactor3000\relax}%
\providecommand \BibitemShut  [1]{\csname bibitem#1\endcsname}%
\let\auto@bib@innerbib\@empty
\bibitem [{\citenamefont {Leggett}(2006)}]{Leggett2006}%
  \BibitemOpen
  \bibfield  {author} {\bibinfo {author} {\bibfnamefont {A.~J.}\ \bibnamefont {Leggett}},\ }\href@noop {} {\emph {\bibinfo {title} {Quantum Liquids: Bose Condensation and Cooper Pairing in Condensed-Matter Systems}}}\ (\bibinfo  {publisher} {Oxford University Press},\ \bibinfo {year} {2006})\BibitemShut {NoStop}%
\bibitem [{\citenamefont {Bloch}\ \emph {et~al.}(2008)\citenamefont {Bloch}, \citenamefont {Dalibard},\ and\ \citenamefont {Zwerger}}]{Bloch2008}%
  \BibitemOpen
  \bibfield  {author} {\bibinfo {author} {\bibfnamefont {I.}~\bibnamefont {Bloch}}, \bibinfo {author} {\bibfnamefont {J.}~\bibnamefont {Dalibard}},\ and\ \bibinfo {author} {\bibfnamefont {W.}~\bibnamefont {Zwerger}},\ }\bibfield  {title} {\bibinfo {title} {Many-body physics with ultracold gases},\ }\href {https://doi.org/10.1103/RevModPhys.80.885} {\bibfield  {journal} {\bibinfo  {journal} {Rev. Mod. Phys.}\ }\textbf {\bibinfo {volume} {80}},\ \bibinfo {pages} {885} (\bibinfo {year} {2008})}\BibitemShut {NoStop}%
\bibitem [{\citenamefont {Ritsch}\ \emph {et~al.}(2013)\citenamefont {Ritsch}, \citenamefont {Domokos}, \citenamefont {Brennecke},\ and\ \citenamefont {Esslinger}}]{Ritsch2013}%
  \BibitemOpen
  \bibfield  {author} {\bibinfo {author} {\bibfnamefont {H.}~\bibnamefont {Ritsch}}, \bibinfo {author} {\bibfnamefont {P.}~\bibnamefont {Domokos}}, \bibinfo {author} {\bibfnamefont {F.}~\bibnamefont {Brennecke}},\ and\ \bibinfo {author} {\bibfnamefont {T.}~\bibnamefont {Esslinger}},\ }\bibfield  {title} {\bibinfo {title} {Cold atoms in cavity-generated dynamical optical potentials},\ }\href {https://doi.org/10.1103/RevModPhys.85.553} {\bibfield  {journal} {\bibinfo  {journal} {Rev. Mod. Phys.}\ }\textbf {\bibinfo {volume} {85}},\ \bibinfo {pages} {553} (\bibinfo {year} {2013})}\BibitemShut {NoStop}%
\bibitem [{\citenamefont {Barreiro}\ \emph {et~al.}(2011)\citenamefont {Barreiro} \emph {et~al.}}]{Barreiro2011}%
  \BibitemOpen
  \bibfield  {author} {\bibinfo {author} {\bibfnamefont {J.~T.}\ \bibnamefont {Barreiro}} \emph {et~al.},\ }\bibfield  {title} {\bibinfo {title} {An open-system quantum simulator with trapped ions},\ }\href {https://doi.org/10.1038/nature09801} {\bibfield  {journal} {\bibinfo  {journal} {Nature}\ }\textbf {\bibinfo {volume} {470}},\ \bibinfo {pages} {486} (\bibinfo {year} {2011})}\BibitemShut {NoStop}%
\bibitem [{\citenamefont {Devoret}\ and\ \citenamefont {Schoelkopf}(2013)}]{Devoret2013}%
  \BibitemOpen
  \bibfield  {author} {\bibinfo {author} {\bibfnamefont {M.~H.}\ \bibnamefont {Devoret}}\ and\ \bibinfo {author} {\bibfnamefont {R.~J.}\ \bibnamefont {Schoelkopf}},\ }\bibfield  {title} {\bibinfo {title} {Superconducting circuits for quantum information: An outlook},\ }\href {https://doi.org/10.1126/science.1231930} {\bibfield  {journal} {\bibinfo  {journal} {Science}\ }\textbf {\bibinfo {volume} {339}},\ \bibinfo {pages} {1169} (\bibinfo {year} {2013})}\BibitemShut {NoStop}%
\bibitem [{\citenamefont {Loss}\ and\ \citenamefont {DiVincenzo}(1998)}]{Loss1998}%
  \BibitemOpen
  \bibfield  {author} {\bibinfo {author} {\bibfnamefont {D.}~\bibnamefont {Loss}}\ and\ \bibinfo {author} {\bibfnamefont {D.~P.}\ \bibnamefont {DiVincenzo}},\ }\bibfield  {title} {\bibinfo {title} {Quantum computation with quantum dots},\ }\href {https://doi.org/10.1103/PhysRevA.57.120} {\bibfield  {journal} {\bibinfo  {journal} {Phys. Rev. A}\ }\textbf {\bibinfo {volume} {57}},\ \bibinfo {pages} {120} (\bibinfo {year} {1998})}\BibitemShut {NoStop}%
\bibitem [{\citenamefont {Cooper}\ \emph {et~al.}(2019)\citenamefont {Cooper}, \citenamefont {Dalibard},\ and\ \citenamefont {Spielman}}]{Cooper2019}%
  \BibitemOpen
  \bibfield  {author} {\bibinfo {author} {\bibfnamefont {N.~R.}\ \bibnamefont {Cooper}}, \bibinfo {author} {\bibfnamefont {J.}~\bibnamefont {Dalibard}},\ and\ \bibinfo {author} {\bibfnamefont {I.~B.}\ \bibnamefont {Spielman}},\ }\bibfield  {title} {\bibinfo {title} {Topological bands for ultracold atoms},\ }\href {https://doi.org/10.1103/RevModPhys.91.015005} {\bibfield  {journal} {\bibinfo  {journal} {Rev. Mod. Phys.}\ }\textbf {\bibinfo {volume} {91}},\ \bibinfo {pages} {015005} (\bibinfo {year} {2019})}\BibitemShut {NoStop}%
\bibitem [{\citenamefont {DiVincenzo}\ and\ \citenamefont {Loss}(1998)}]{DiVincenzo1998}%
  \BibitemOpen
  \bibfield  {author} {\bibinfo {author} {\bibfnamefont {D.~P.}\ \bibnamefont {DiVincenzo}}\ and\ \bibinfo {author} {\bibfnamefont {D.}~\bibnamefont {Loss}},\ }\bibfield  {title} {\bibinfo {title} {Quantum computation with quantum dots},\ }\href {https://doi.org/10.1103/PhysRevA.57.120} {\bibfield  {journal} {\bibinfo  {journal} {Phys. Rev. A}\ }\textbf {\bibinfo {volume} {57}},\ \bibinfo {pages} {120} (\bibinfo {year} {1998})}\BibitemShut {NoStop}%
\bibitem [{\citenamefont {Lin}\ \emph {et~al.}(2009)\citenamefont {Lin} \emph {et~al.}}]{Lin2009}%
  \BibitemOpen
  \bibfield  {author} {\bibinfo {author} {\bibfnamefont {Y.-J.}\ \bibnamefont {Lin}} \emph {et~al.},\ }\bibfield  {title} {\bibinfo {title} {Synthetic magnetic fields for ultracold neutral atoms},\ }\href {https://doi.org/10.1038/nature08609} {\bibfield  {journal} {\bibinfo  {journal} {Nature}\ }\textbf {\bibinfo {volume} {462}},\ \bibinfo {pages} {628} (\bibinfo {year} {2009})}\BibitemShut {NoStop}%
\bibitem [{\citenamefont {Chang}\ \emph {et~al.}(2004)\citenamefont {Chang} \emph {et~al.}}]{Chang2004}%
  \BibitemOpen
  \bibfield  {author} {\bibinfo {author} {\bibfnamefont {M.~S.}\ \bibnamefont {Chang}} \emph {et~al.},\ }\bibfield  {title} {\bibinfo {title} {Observation of spinor dynamics in optically trapped {87Rb} {Bose-Einstein} condensates},\ }\href {https://doi.org/10.1103/PhysRevLett.92.140403} {\bibfield  {journal} {\bibinfo  {journal} {Phys. Rev. Lett.}\ }\textbf {\bibinfo {volume} {92}},\ \bibinfo {pages} {140403} (\bibinfo {year} {2004})}\BibitemShut {NoStop}%
\bibitem [{\citenamefont {Matuszewski}(2010)}]{Matuszewski2010}%
  \BibitemOpen
  \bibfield  {author} {\bibinfo {author} {\bibfnamefont {M.}~\bibnamefont {Matuszewski}},\ }\bibfield  {title} {\bibinfo {title} {Ground states of trapped spin-1 condensates in magnetic field},\ }\href {https://doi.org/10.1103/PhysRevA.82.053630} {\bibfield  {journal} {\bibinfo  {journal} {Phys. Rev. A}\ }\textbf {\bibinfo {volume} {82}},\ \bibinfo {pages} {053630} (\bibinfo {year} {2010})}\BibitemShut {NoStop}%
\bibitem [{\citenamefont {Horne}(2007)}]{Horne2007}%
  \BibitemOpen
  \bibfield  {author} {\bibinfo {author} {\bibfnamefont {R.~A.}\ \bibnamefont {Horne}},\ }\bibfield  {title} {\bibinfo {title} {A cylindrically symmetric magnetic trap for {Bose-Einstein} condensate atom interferometry applications},\ }\href@noop {} {\bibfield  {journal} {\bibinfo  {journal} {Thesis, University of Virginia}\ } (\bibinfo {year} {2007})}\BibitemShut {NoStop}%
\bibitem [{\citenamefont {Chin}\ \emph {et~al.}(2010)\citenamefont {Chin}, \citenamefont {Grimm}, \citenamefont {Julienne},\ and\ \citenamefont {Tiesinga}}]{Chin2010}%
  \BibitemOpen
  \bibfield  {author} {\bibinfo {author} {\bibfnamefont {C.}~\bibnamefont {Chin}}, \bibinfo {author} {\bibfnamefont {R.}~\bibnamefont {Grimm}}, \bibinfo {author} {\bibfnamefont {P.}~\bibnamefont {Julienne}},\ and\ \bibinfo {author} {\bibfnamefont {E.}~\bibnamefont {Tiesinga}},\ }\bibfield  {title} {\bibinfo {title} {Feshbach resonances in ultracold gases},\ }\href {https://doi.org/10.1103/RevModPhys.82.1225} {\bibfield  {journal} {\bibinfo  {journal} {Rev. Mod. Phys.}\ }\textbf {\bibinfo {volume} {82}},\ \bibinfo {pages} {1225} (\bibinfo {year} {2010})}\BibitemShut {NoStop}%
\bibitem [{\citenamefont {Petrich}\ \emph {et~al.}(1995)\citenamefont {Petrich}, \citenamefont {Anderson}, \citenamefont {Ensher},\ and\ \citenamefont {Cornell}}]{Petrich1995}%
  \BibitemOpen
  \bibfield  {author} {\bibinfo {author} {\bibfnamefont {W.}~\bibnamefont {Petrich}}, \bibinfo {author} {\bibfnamefont {M.~H.}\ \bibnamefont {Anderson}}, \bibinfo {author} {\bibfnamefont {J.~R.}\ \bibnamefont {Ensher}},\ and\ \bibinfo {author} {\bibfnamefont {E.~A.}\ \bibnamefont {Cornell}},\ }\bibfield  {title} {\bibinfo {title} {Stable, tightly confining magnetic trap for evaporative cooling of neutral atoms},\ }\href {https://doi.org/10.1103/PhysRevLett.74.3352} {\bibfield  {journal} {\bibinfo  {journal} {Phys. Rev. Lett.}\ }\textbf {\bibinfo {volume} {74}},\ \bibinfo {pages} {3352} (\bibinfo {year} {1995})}\BibitemShut {NoStop}%
\bibitem [{\citenamefont {Horne}\ and\ \citenamefont {Sackett}(2017)}]{Horne2017}%
  \BibitemOpen
  \bibfield  {author} {\bibinfo {author} {\bibfnamefont {R.~A.}\ \bibnamefont {Horne}}\ and\ \bibinfo {author} {\bibfnamefont {C.~A.}\ \bibnamefont {Sackett}},\ }\bibfield  {title} {\bibinfo {title} {A cylindrically symmetric magnetic trap for compact {Bose-Einstein} condensate atom interferometer gyroscopes},\ }\href {https://doi.org/10.1063/1.4973123} {\bibfield  {journal} {\bibinfo  {journal} {Rev. Sci. Instrum.}\ }\textbf {\bibinfo {volume} {88}},\ \bibinfo {pages} {023101} (\bibinfo {year} {2017})}\BibitemShut {NoStop}%
\bibitem [{\citenamefont {Pezzè}\ \emph {et~al.}(2018)\citenamefont {Pezzè} \emph {et~al.}}]{Pezze2018}%
  \BibitemOpen
  \bibfield  {author} {\bibinfo {author} {\bibfnamefont {L.}~\bibnamefont {Pezzè}} \emph {et~al.},\ }\bibfield  {title} {\bibinfo {title} {Quantum metrology with nonclassical states of atomic ensembles},\ }\href {https://doi.org/10.1103/RevModPhys.90.035005} {\bibfield  {journal} {\bibinfo  {journal} {Rev. Mod. Phys.}\ }\textbf {\bibinfo {volume} {90}},\ \bibinfo {pages} {035005} (\bibinfo {year} {2018})}\BibitemShut {NoStop}%
\bibitem [{\citenamefont {Livi}\ \emph {et~al.}(2016)\citenamefont {Livi} \emph {et~al.}}]{PhysRevLett.117.220401}%
  \BibitemOpen
  \bibfield  {author} {\bibinfo {author} {\bibfnamefont {L.~F.}\ \bibnamefont {Livi}} \emph {et~al.},\ }\bibfield  {title} {\bibinfo {title} {Synthetic dimensions and spin-orbit coupling with an optical clock transition},\ }\href {https://doi.org/10.1103/PhysRevLett.117.220401} {\bibfield  {journal} {\bibinfo  {journal} {Phys. Rev. Lett.}\ }\textbf {\bibinfo {volume} {117}},\ \bibinfo {pages} {220401} (\bibinfo {year} {2016})}\BibitemShut {NoStop}%
\bibitem [{\citenamefont {Lundblad}\ \emph {et~al.}(2019)\citenamefont {Lundblad}, \citenamefont {Carollo}, \citenamefont {Lannert} \emph {et~al.}}]{Lundblad2019}%
  \BibitemOpen
  \bibfield  {author} {\bibinfo {author} {\bibfnamefont {N.}~\bibnamefont {Lundblad}}, \bibinfo {author} {\bibfnamefont {R.~A.}\ \bibnamefont {Carollo}}, \bibinfo {author} {\bibfnamefont {C.}~\bibnamefont {Lannert}}, \emph {et~al.},\ }\bibfield  {title} {\bibinfo {title} {Shell potentials for microgravity {Bose-Einstein} condensates},\ }\href {https://doi.org/10.1038/s41526-019-0087-y} {\bibfield  {journal} {\bibinfo  {journal} {npj Microgravity}\ }\textbf {\bibinfo {volume} {5}},\ \bibinfo {pages} {30} (\bibinfo {year} {2019})}\BibitemShut {NoStop}%
\bibitem [{\citenamefont {Tononi}\ and\ \citenamefont {Salasnich}(2019)}]{PhysRevLett.123.160403}%
  \BibitemOpen
  \bibfield  {author} {\bibinfo {author} {\bibfnamefont {A.}~\bibnamefont {Tononi}}\ and\ \bibinfo {author} {\bibfnamefont {L.}~\bibnamefont {Salasnich}},\ }\bibfield  {title} {\bibinfo {title} {Bose-einstein condensation on the surface of a sphere},\ }\href {https://doi.org/10.1103/PhysRevLett.123.160403} {\bibfield  {journal} {\bibinfo  {journal} {Phys. Rev. Lett.}\ }\textbf {\bibinfo {volume} {123}},\ \bibinfo {pages} {160403} (\bibinfo {year} {2019})}\BibitemShut {NoStop}%
\bibitem [{\citenamefont {Grimm}\ \emph {et~al.}(2000)\citenamefont {Grimm}, \citenamefont {Weidemüller},\ and\ \citenamefont {Ovchinnikov}}]{GRIMM200095}%
  \BibitemOpen
  \bibfield  {author} {\bibinfo {author} {\bibfnamefont {R.}~\bibnamefont {Grimm}}, \bibinfo {author} {\bibfnamefont {M.}~\bibnamefont {Weidemüller}},\ and\ \bibinfo {author} {\bibfnamefont {Y.~B.}\ \bibnamefont {Ovchinnikov}},\ }\bibfield  {title} {\bibinfo {title} {Optical dipole traps for neutral atoms},\ }\href@noop {} {\bibfield  {journal} {\bibinfo  {journal} {Adv. Atom. Mol. Opt. Phy.}\ }\textbf {\bibinfo {volume} {42}},\ \bibinfo {pages} {95} (\bibinfo {year} {2000})}\BibitemShut {NoStop}%
\bibitem [{\citenamefont {Luiten}\ \emph {et~al.}(1996)\citenamefont {Luiten}, \citenamefont {Reynolds},\ and\ \citenamefont {Walraven}}]{PhysRevA.53.381}%
  \BibitemOpen
  \bibfield  {author} {\bibinfo {author} {\bibfnamefont {O.~J.}\ \bibnamefont {Luiten}}, \bibinfo {author} {\bibfnamefont {M.~W.}\ \bibnamefont {Reynolds}},\ and\ \bibinfo {author} {\bibfnamefont {J.~T.~M.}\ \bibnamefont {Walraven}},\ }\bibfield  {title} {\bibinfo {title} {Kinetic theory of the evaporative cooling of a trapped gas},\ }\href {https://doi.org/10.1103/PhysRevA.53.381} {\bibfield  {journal} {\bibinfo  {journal} {Phys. Rev. A}\ }\textbf {\bibinfo {volume} {53}},\ \bibinfo {pages} {381} (\bibinfo {year} {1996})}\BibitemShut {NoStop}%
\bibitem [{\citenamefont {Berg-S\o{}rensen}(1997)}]{PhysRevA.55.1281}%
  \BibitemOpen
  \bibfield  {author} {\bibinfo {author} {\bibfnamefont {K.}~\bibnamefont {Berg-S\o{}rensen}},\ }\bibfield  {title} {\bibinfo {title} {Kinetics for evaporative cooling of a trapped gas},\ }\href {https://doi.org/10.1103/PhysRevA.55.1281} {\bibfield  {journal} {\bibinfo  {journal} {Phys. Rev. A}\ }\textbf {\bibinfo {volume} {55}},\ \bibinfo {pages} {1281} (\bibinfo {year} {1997})}\BibitemShut {NoStop}%
\bibitem [{\citenamefont {Ketterle}\ and\ \citenamefont {Druten}(1996)}]{WOS:A1996BH82Z00004}%
  \BibitemOpen
  \bibfield  {author} {\bibinfo {author} {\bibfnamefont {W.}~\bibnamefont {Ketterle}}\ and\ \bibinfo {author} {\bibfnamefont {N.~V.}\ \bibnamefont {Druten}},\ }\bibfield  {title} {\bibinfo {title} {Evaporative cooling of trapped atoms},\ }\href@noop {} {\bibfield  {journal} {\bibinfo  {journal} {Adv. Atom. Mol. Opt. Phy.}\ }\textbf {\bibinfo {volume} {37}},\ \bibinfo {pages} {181} (\bibinfo {year} {1996})}\BibitemShut {NoStop}%
\bibitem [{\citenamefont {Moan}\ \emph {et~al.}(2020)\citenamefont {Moan}, \citenamefont {Horne}, \citenamefont {Arpornthip}, \citenamefont {Luo}, \citenamefont {Fallon}, \citenamefont {Berl},\ and\ \citenamefont {Sackett}}]{Moan2020}%
  \BibitemOpen
  \bibfield  {author} {\bibinfo {author} {\bibfnamefont {E.~R.}\ \bibnamefont {Moan}}, \bibinfo {author} {\bibfnamefont {R.~A.}\ \bibnamefont {Horne}}, \bibinfo {author} {\bibfnamefont {T.}~\bibnamefont {Arpornthip}}, \bibinfo {author} {\bibfnamefont {Z.}~\bibnamefont {Luo}}, \bibinfo {author} {\bibfnamefont {A.~J.}\ \bibnamefont {Fallon}}, \bibinfo {author} {\bibfnamefont {S.~J.}\ \bibnamefont {Berl}},\ and\ \bibinfo {author} {\bibfnamefont {C.~A.}\ \bibnamefont {Sackett}},\ }\bibfield  {title} {\bibinfo {title} {Quantum rotation sensing with dual sagnac interferometers in an atom-optical waveguide},\ }\href {https://doi.org/10.1103/PhysRevLett.124.120403} {\bibfield  {journal} {\bibinfo  {journal} {Phys. Rev. Lett.}\ }\textbf {\bibinfo {volume} {124}},\ \bibinfo {pages} {120403} (\bibinfo {year} {2020})}\BibitemShut {NoStop}%
\bibitem [{\citenamefont {Beydler}\ \emph {et~al.}(2024)\citenamefont {Beydler}, \citenamefont {Moan}, \citenamefont {Luo}, \citenamefont {Chu},\ and\ \citenamefont {Sackett}}]{Beydler2023}%
  \BibitemOpen
  \bibfield  {author} {\bibinfo {author} {\bibfnamefont {M.~M.}\ \bibnamefont {Beydler}}, \bibinfo {author} {\bibfnamefont {E.~R.}\ \bibnamefont {Moan}}, \bibinfo {author} {\bibfnamefont {Z.}~\bibnamefont {Luo}}, \bibinfo {author} {\bibfnamefont {Z.}~\bibnamefont {Chu}},\ and\ \bibinfo {author} {\bibfnamefont {C.~A.}\ \bibnamefont {Sackett}},\ }\bibfield  {title} {\bibinfo {title} {Guided-wave sagnac atom interferometer with large area and multiple orbits},\ }\href {https://doi.org/10.1116/5.0173769} {\bibfield  {journal} {\bibinfo  {journal} {AVS Quantum Sci.}\ }\textbf {\bibinfo {volume} {6}},\ \bibinfo {pages} {014401} (\bibinfo {year} {2024})}\BibitemShut {NoStop}%
\end{thebibliography}%

\end{document}